\documentclass[prb,aps,twocolumn,showpacs,groupedaddress,superscriptaddress]{revtex4}

\usepackage{graphicx,epsf,bm,bbm,color,amsmath,ulem,amssymb}

  \newcommand{\tr}[1]{\textrm{#1}}

	\newcommand{\beq}{\begin{eqnarray}}
	
	\newcommand{\eeq}{\end{eqnarray}}
	\newcommand{\nn}{\nonumber}

\newcommand{\bem}{\begin{pmatrix}}
\newcommand{\eem}{\end{pmatrix}}

\newcommand{\f}{\frac}

\begin{document}

\title{Quantum Oscillations in Nodal Line Systems} 
 \author{Hui Yang}
 \affiliation{International Center for Quantum Materials, School of Physics, Peking University, Beijing 100871, P. R. China}
\author{Roderich Moessner}
\affiliation{Max-Planck-Institut f\"ur Physik komplexer Systeme, D-01187 Dresden, Germany}
\author{Lih-King Lim}\email{lihking@123mail.org}
\affiliation{Zhejiang Institute of Modern Physics, Zhejiang University, Hangzhou 310027, P. R. China}
 \affiliation{Institute for Advanced Study, Tsinghua University, Beijing 100084, P. R. China}

\date{\today}

\begin{abstract}
We study signatures of magnetic quantum oscillations in three-dimensional nodal line semimetals at zero temperature. The extended nature of the degenerate bands can result in a Fermi surface geometry with topological genus one, as well as a Fermi surface of electron and hole pockets encapsulating the nodal line. Moreover, the underlying two-band model to describe a nodal line is not unique, in that there are two classes of Hamiltonian with distinct band topology giving rise to the same Fermi surface geometry. After identifying the extremal cyclotron orbits in various magnetic field directions, we study their concomitant Landau levels and resulting quantum oscillation signatures. By Landau-fan-diagram analyses we extract the non-trivial $\pi$ Berry phase signature for extremal orbits linking the nodal line.  
\end{abstract}
\maketitle

\section{Introduction}
Nodal line systems are a new kind of three-dimensional topological semimetal of great current interests \cite{Burkov11b,Phillips14,Weng15b,Fang15,Mullen15,Kim15,Yu15,Chen15,Heikkila15,Xie15,Chan15,Bian16,Yamakage16,Ezawa16,Wang16,RLi16,Yan16,Okamoto16,Lim17,Hirayama17,Okugawa17}. They extend the concept of  Weyl semimetal \cite{Murakami07,Wan11,Weng15a,Xu15,Lu15,Lv15a,Lv15b} with its point-like crossing between conduction and valence bands to a band touching that forms a closed loop \cite{Burkov11b,Phillips14,Nand16,Roy17,Kim17}. The stability condition for this kind of degeneracy (or to ensure `gaplessness') turns out to be easily met, with various proposals for such types of systems now available \cite{Weng15b,Mullen15,Kim15,RLi16,Hirayama17,Jin17}. Moreover, they can be realized in materials with \cite{Kim15,Ezawa16} or without \cite{Weng15b,Fang15,Mullen15,Yu15,Chen15,Wang16,RLi16,Okugawa17} spin-orbit couplings. Indeed many recent experimental studies have shown promising progress towards its realization \cite{Wu16,Schoop16,Neupane16,Hu16}.

The extended nature of the closed loop band touching offers new perspectives for studies of topological physics. First, the band touching line need not be a constant energy contour. This results in an elaborate Fermi surface geometry typically with electron and hole pockets encapsulating the nodal line \cite{Kim15, Jin17, Hyart17}. Second, by viewing the band touching as a pseudospin structure in momentum space \cite{Volovik03}, there correspond different kinds of topological defect, arising from distinct Hamiltonians \cite{Kim15, Lim17}, associated with the same nodal line energy spectrum. This is not the case for a Weyl semimetal \cite{Volovik03, Murakami07} where a pseudospin monopole is uniquely associated with the point-like defect. On the other hand, the nodal line requires only a topological $\pi$ Berry phase in loops encircling the nodal line \cite{Kim15}. In Ref. \cite{Lim17}, we studied a realization of vortex ring pseudospin defect, in contrast to the ordinary nodal line Hamiltonian \cite{Kim15}, which exhibits a `maximal' three-dimensional anomalous Hall effect. Given these rich nodal line systems, we study in this paper their magnetic transport signatures in a few characteristic settings.

We undertake a theoretical study of the characterization of various nodal lines in terms of quantum oscillations in the presence of a magnetic field at zero tempertaure \cite{Ashcroft,Shoenberg}. Specifically, we study nodal lines distinguished by two distinct kinds of pseudospin defects, i.e., the ordinary \cite{Kim15} and the vortex ring \cite{Lim17} Hamiltonians. We consider cases where the nodal line does and does not lie on a constant energy contour. We then study the resulting quantum oscillation signatures for magnetic fields along various symmetry axes of the nodal line.

As we shall see, due to the elaborate Fermi surface geometry, there can be multiple extremal cyclotron orbits, which can even intersect each other, in the presence of a magnetic field. At first sight this raises some ambiguities in the identification of oscillation frequencies with the momentum space area covered using simple semi-classic arguments. We therefore first focus on obtaining the LLs (either analytically or numerically) and then study the induced frequencies to be identified with the corresponding cyclotron orbits.
Finally, LLs also carry important topological information of the Hamiltonians, namely the Berry phase picked up by the orbit. These are extracted by Landau-fan-diagram analyses. 

The paper is organized as follows. In Sect. II, we describe the Fermi surface geometry given the various types of nodal line Hamiltonians and identify the extremal orbits in the presence of a magnetic field. And we summarize the procedures to relate the Landau levels with the quantum oscillation periods and the Berry phases. In Sect. III and Sect. IV, we study the associated quantum oscillations for cases of equi-energy and tilted-energy nodal lines, respectively. We end with conclusions and an outlook in Sect. V.

\section{Fermi surface geometry, extremal orbits and Landau levels}
\subsection{Fermi surface geometry}
An equi-energy nodal line Hamiltonian \cite{Kim15,Fang15,Mullen15} is given  by
\beq
H(p_x,p_y,p_z) =\biggr(   \f{1}{2m_r} (p_x^2+p_y^2-p_0^2) \biggr)\sigma_x+ v_z p_z  \sigma_y,
\eeq
where $\bm{\sigma}$ are Pauli matrices acting on orbital/sublattice space, $m_r$ gives the band mass on the $(p_x,p_y)$ plane and $v_z$ the speed in the $p_z$ direction. The upper and lower energy bands touch on a circle, $p_x^2+p_y^2=p_0^2$, $p_z=0$, with radius $p_0$ at constant energy $E=0$. For a Fermi energy slightly larger than $E=0$, the Fermi surface of electrons (or holes for $E<0$) takes the shape of a torus - of genus 1 topology - which encloses the degenerate closed line, see Fig. 1a. As the energy is further increased, the volume of the Fermi surface increases whereupon a critical volume is reached, beyond which its geometry changes to a `sphere' of genus 0. 

As two of us studied in Ref.~\cite{Lim17}, the Hamiltonian giving rise to a nodal line is not unique. In the absence of time-reversal and inversion symmetries, the same nodal line ($p_x^2+p_y^2=p_0^2$, $p_z=0$ with $E=0$) can also result from the Hamiltonian 
\beq
H_{\textrm{vx}}(p_x,p_y,p_z) &=&-\f{1}{m_\bot}p_x p_z \sigma_x-\f{1}{m_\bot} p_y p_z \sigma_y\nn\\
&&\!\!\!\!\!\!\!\!\!\! +\biggl( \f{1}{2m_r} (p_x^2+p_y^2-p_z^2)  -\f{p_0^2}{2 m_r} \biggr)\sigma_z,
\eeq 
which exhibits a pseudospin vortex ring defect structure in momentum space. In comparison to $H$, despite sharing the same Fermi surface geometry, the pseudospin vortex ring defect leads to a maximal anomalous Hall effect \cite{Lim17}.  

\begin{figure}
\begin{center}
\includegraphics[width=8.9cm]{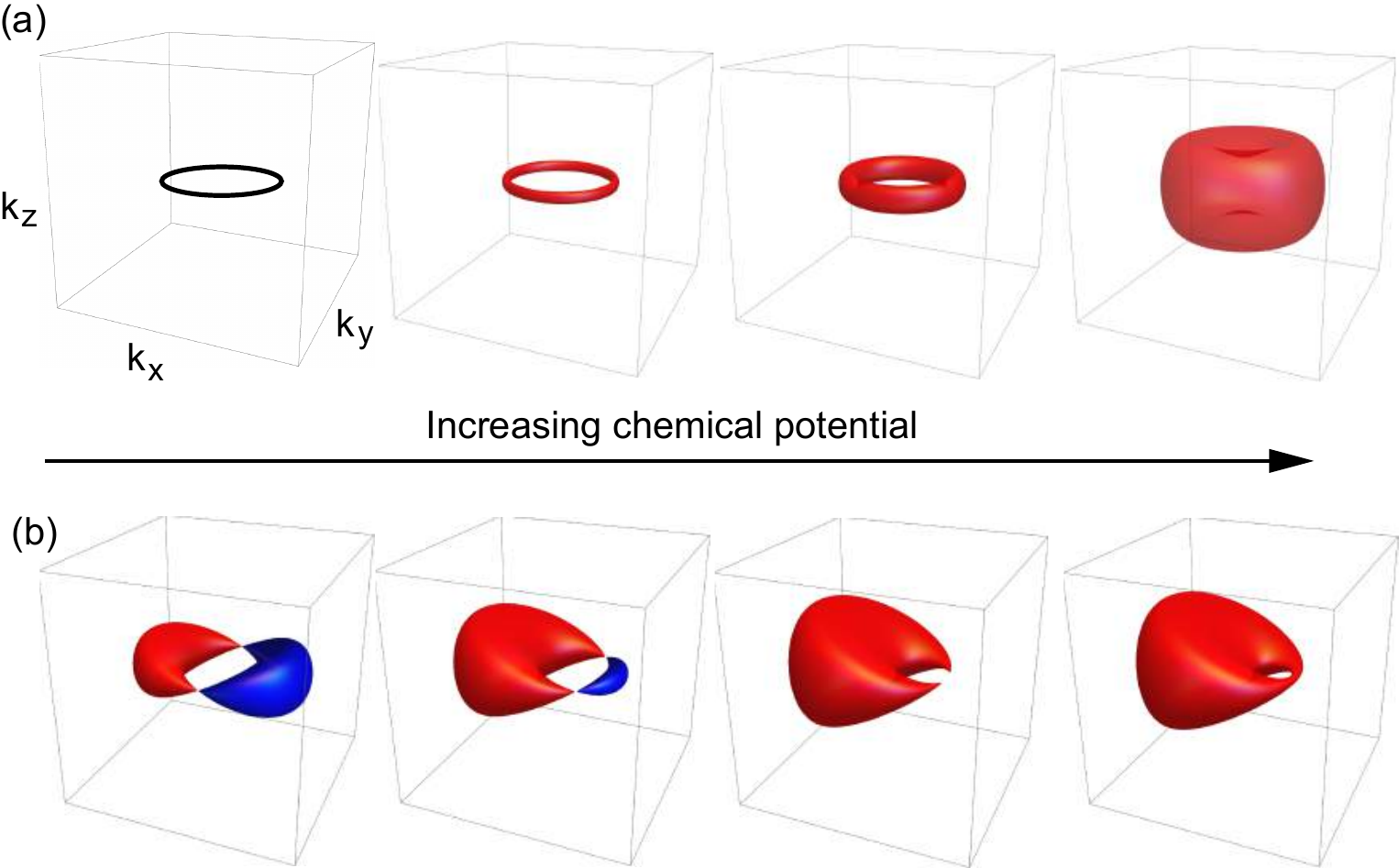}
\end{center}
\caption{Fermi surface of (a) equi-energy nodal line; (b) tilted-energy nodal line. }
\end{figure}

In addition to different pesudospin textures, a nodal line need not be restricted to occur at the same energy ($E=0$) \cite{Kim15}. Most simply, 
\beq 
H_{\textrm{tilt}}=v p_x\,\mathbb{I} +H,
\eeq
where the identity term introduces an energy slope (`tilt') in the $p_x$ direction (similarly for $H_{\tr{vx}}$). The Fermi surface at $E=0$ now consists of an electron and a hole pocket touching at two points, see Fig.~1b. As the Fermi energy increases, the two touching points move towards each other along the underlying nodal line with growing (shrinking) electron (hole) pocket. At a critical Fermi energy, the hole pocket vanishes, leaving behind only an electron Fermi surface of genus 1. Beyond, the Fermi surface evolution with increasing energy displays a similar behavior to the equi-energy nodal line case. 

In this paper, we consider two characteristic scenarios for ease of study. First, the band touching loop is assumed to be a perfect circle. In actual realizations, it is more likely to be a wire-loop of arbitrary shape in momentum space. Second, we introduce a simple form of energy inhomogeneity (by tilting the normal of the nodal circle plane with respect to the energy gradient) resulting in one electron/hole pocket pair. A more complex energy landscape can lead to multiple electron/hole pockets. Nevertheless, the following results can be generalized to these cases.

\subsection{Extremal orbits}
We consider two characteristic magnetic $B$-field directions in turn: perpendicular and parallel to the nodal line plane ($p_z=0$ plane). For the cyclotron orbits, we specify electron momenta which give extremal areas on the Fermi surface as they dominate quantum oscillation signatures \cite{Ashcroft,Shoenberg}.

\begin{figure}
\begin{center}
\includegraphics[width=7.5cm]{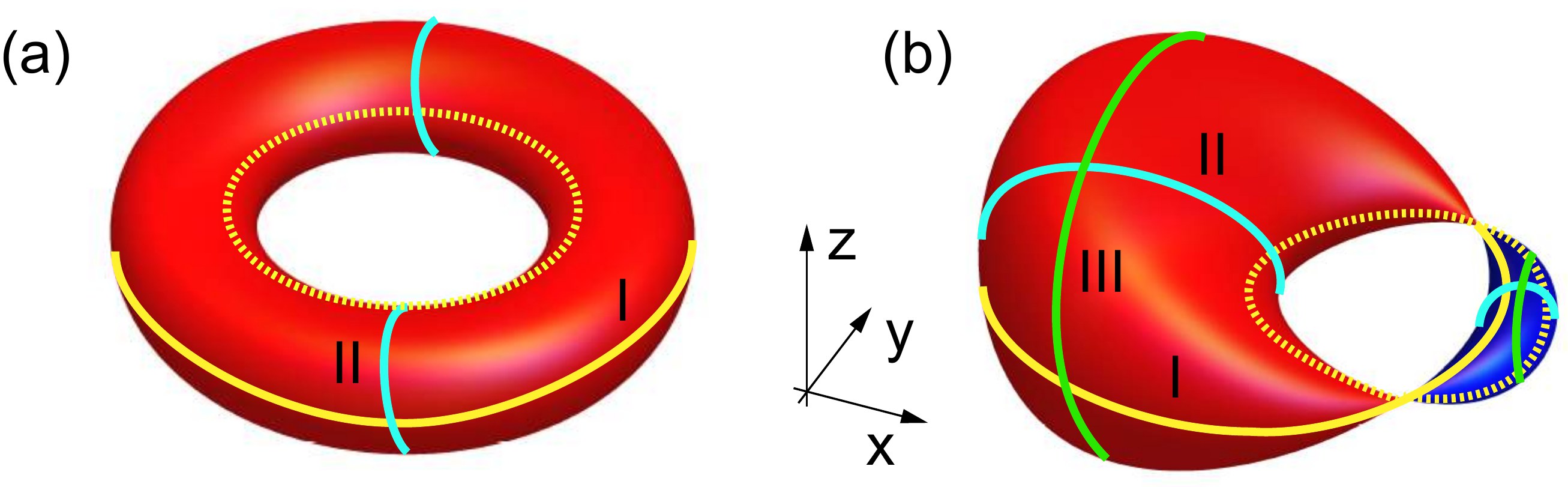}
\end{center}
\caption{Semiclassic extremal orbits (labelled as I, II and III) for (a) the equi-energy and (b) tilted-energy nodal lines for various magnetic field directions.}
\end{figure} 
\subsubsection*{Perpendicular magnetic field}
For a magnetic field perpendicular to the nodal line plane, the extremal cyclotron orbits are formed by a section on the $p_z=0$ plane. 

For the equi-energy nodal line at finite Fermi energy, there are two such orbits each encircling the outer and the inner side of the doughnut surface, respectively (orbit I in Fig. 2a). They are traversed in opposite directions. 

For the tilted-energy nodal line, there are also two extremal circular orbits following the periphery at $p_z=0$ plane, which intersect each other at two points where the electron and hole pockets meet (for zero Fermi energy) (orbit I in Fig.~2b). This can be inferred by comparing the orbit area for different sectional cuts of the Fermi surface with nearby $p_z$ values. For finite Fermi energy, the extremal circular orbits remain at $p_z=0$ while their centers move relatively towards each other. At the critical Fermi energy, the intersect points meet at one point with one orbit lying entirely within the other. At higher Fermi energy, the inner and outer orbits resemble the situation of the equi-energy nodal line. 

\subsubsection*{Parallel magnetic field}
For the equi-energy nodal line and below the critical Fermi energy, a magnetic field in parallel to the nodal circle plane results in two orbits on the perpendicular section of the toroidal Fermi surface (orbit II in Fig.~2a). They cover the same area and the same direction of motion. Above the critical Fermi energy, the two orbits join to form a single closed orbit.

For the tilted-energy nodal line, we further distinguish two field directions along the high symmetry axes, namely in the $y$- and $x$-directions (orbits II and III, respectively, in Fig.~2b). For these cases, there are generically two extremal orbits, encircling the electron and hole pockets respectively, below the critical Fermi energy. The ways in which they encircle the underlying nodal line are different and cover different areas, leading to different physical consequences.

\subsubsection*{Quantum oscillations in a magnetic field}
Having specified the geometry of the Fermi surface and the extremal orbits, we outline the procedures for the study of the associated magnetic quantum oscillation signatures following Onsager's relation \cite{Onsager52}. 

First, with the LLs associated with the extremal orbits, we determine a discrete set of magnetic field values $\{B_n(\mu)\}$, indexed by integer-valued $n$, at which the LLs intersect with a fixed \cite{ft1} chemical potential $\mu$. The set obeys Onsager's relation: $S(\mu)=(n+\gamma)\,(2 \pi e/\hbar)\, B_n(\mu)$, where $S(\mu)$ is the $k$-space area enclosed by the extremal cyclotron orbit on the Fermi surface, and $0\leq\gamma<1$ is related to the Berry phase of that orbit \cite{Novoselov05, Gusynin05, Zhang05, Mikitik99, Taskin11, Fuchs10}. Specifically, the fundamental sectional area $S_{FO}$ corresponding to a particular closed orbit (giving the fundamental oscillation (FO) frequency) is given by $(1/S_{FO})=((1/B_{n})-(1/B_{n'}))(\hbar/2\pi e)$.
The last expression is obtained by eliminating $\gamma$ in Onsager's relation with two magnetic field values belonging to the same cyclotron orbit with unit index difference (i.e., $|n-n'|=1$). Therefore, because there typically exists more than one extremal orbit in a realistic bandstructure, the choice in the grouping of $B_n$'s belonging to the same set has to be supplemented with some minimal knowledge about the Fermi surface and the resulting cyclotron orbits.

Second, corresponding to the set $\{ B_n \}$, we make the $1/B_n$ versus $n$ plot. The slope of the interpolating line gives the inverse cross-sectional area $1/S_{FO}$ and by changing the chemical potential gradually without crossing the critical Fermi energy, lines of different slopes (corresponding to gradually changing the cross-section area) form the Landau fan diagram. The abscissa (the intercept on the $n$ axis) of the fan diagram gives $-\gamma$. Typically, a value of $\gamma=0 \,(1/2)$ indicates (no) topological Berry phase of $\pi$ \cite{Novoselov05, Gusynin05, Zhang05, Mikitik99, Taskin11, Fuchs10}.

We now study the oscillations and Berry phase signatures as derived from Landau level analyses. In the following, where an analytic LL expression is absent we extract its value with the numerical Landau level structure. Sec. III is devoted to the equi-energy nodal line and Sec. IV is devoted to the tilted-energy nodal line. 

\begin{figure}
\begin{center}
\includegraphics[width=8.7cm]{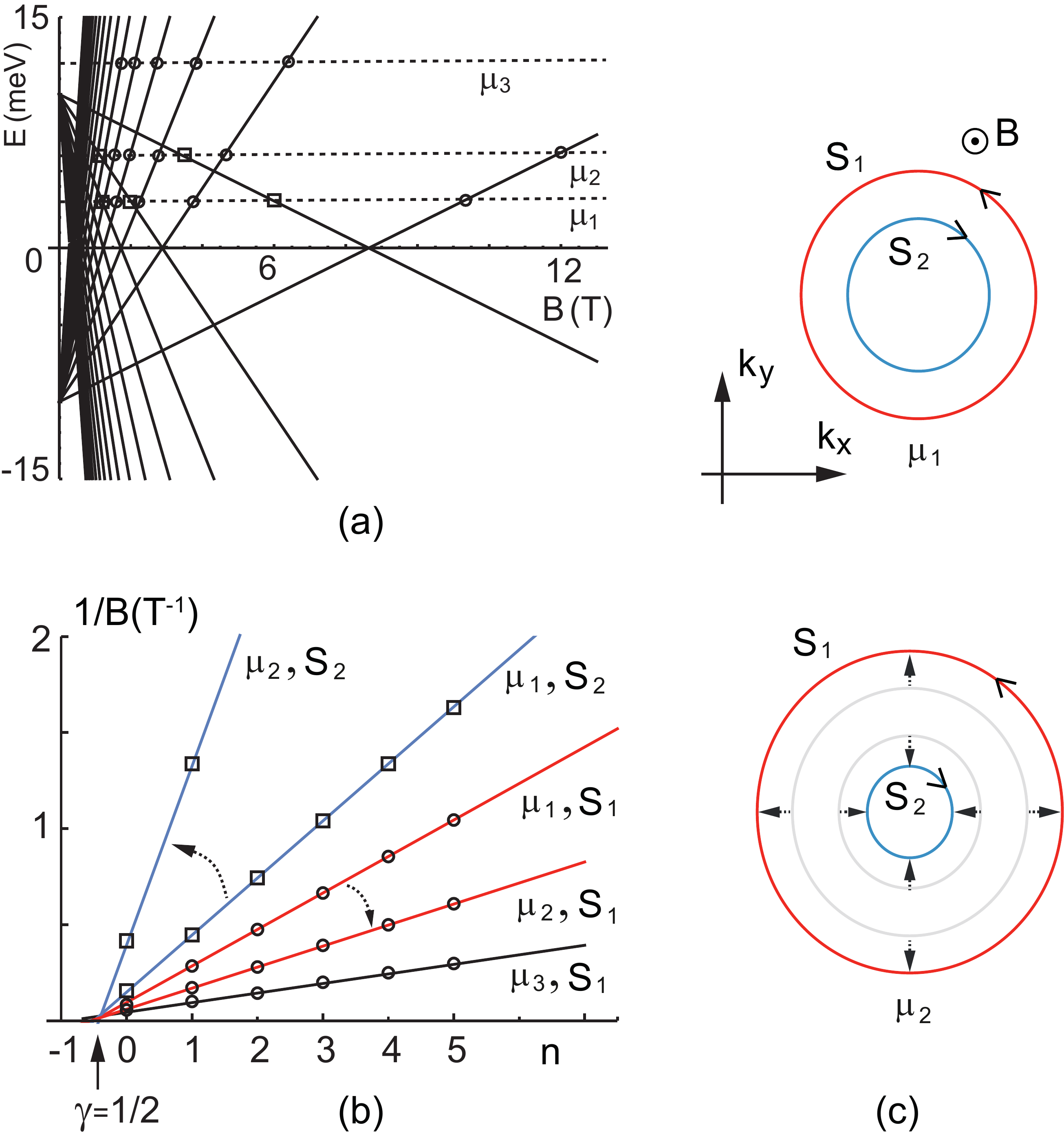}
\end{center}
\caption{(a) Perpendicular field Landau levels (full lines) of the equi-energy nodal line\cite{ft2} at $p_z=0$. Crossings with a fixed chemical potential (circles and squares) form the two sets $\{B_{n1,n2}\,({\mu})\}$. (b) Landau fan diagram with three chemical potential values $\mu_{1-3}$ and the two corresponding cross sectional areas $S_{1,2}$. (c) Extremal orbits for $\mu_{1,2}$. The arrows on the orbits indicate the cyclotron motion direction. The dashed arrows indicate the direction of increasing chemical potential.}
\end{figure}
\section{Landau level results for equi-energy nodal line}
We study the two topologically inequivalent equi-energy nodal line models. While the two Hamiltonians give the same Fermi surface geometry, we can distinguish them by their LL structure in the parallel field case.

\subsection{Perpendicular field direction (orbit I)}
The extremal orbits lie at momentum $p_z=0$ where the two kinds of nodal line Hamiltonian reduce to the same Hamiltonian $H(p_x,p_y,0)=H_{\textrm{vx}}(p_x,p_y,0)$. Their LLs are given by: 
\beq
E_n=\pm\left((n+\frac{1}{2})\epsilon_B-\Delta\right),
\eeq
$n=0,1,2, \dots$, a result of quantizing the energy spectrum of two parabolae inverted with respect to each other (and shifted by a finite energy difference) with the associated energy scale $\epsilon_B\equiv e B\hbar/m_r$. For $\mu<\Delta$, we identify the two sets $\{B_{n1}(\mu)\}$, $\{B_{n2}(\mu)\}$ with LL of opposite slopes (marked with open circles and squares, respectively), see Fig.~3a. By computing the cross sectional areas $S_1$ and $S_2$, they can be identified with the outer and inner orbits, respectively, on the peripheries of the genus 1 Fermi surface (see orbit I in Fig.~2a and Fig.~3c). They give the two fundamental frequencies of quantum oscillations. 

As $\mu$ increases, the outer (inner) orbit increases (shrinks) (Fig.~3c). In the fan diagram they correspond to decreasing (increasing) line slopes (Fig.~3b). They show a common intercept, giving $\gamma=1/2$, indicating no Berry phase, as expected for orbits circulating a parabolic band with no Berry phase. For $\mu>\Delta$ ($\mu_3$ in Fig.~3a), the inner orbit diminishes and only one set of LLs of positive slope remains, corresponding to the single outer orbit circulating the Fermi surface of genus 0. Thus, only one fundamental oscillation frequency remains. 

\begin{figure}
\begin{center}
\includegraphics[width=8.7cm]{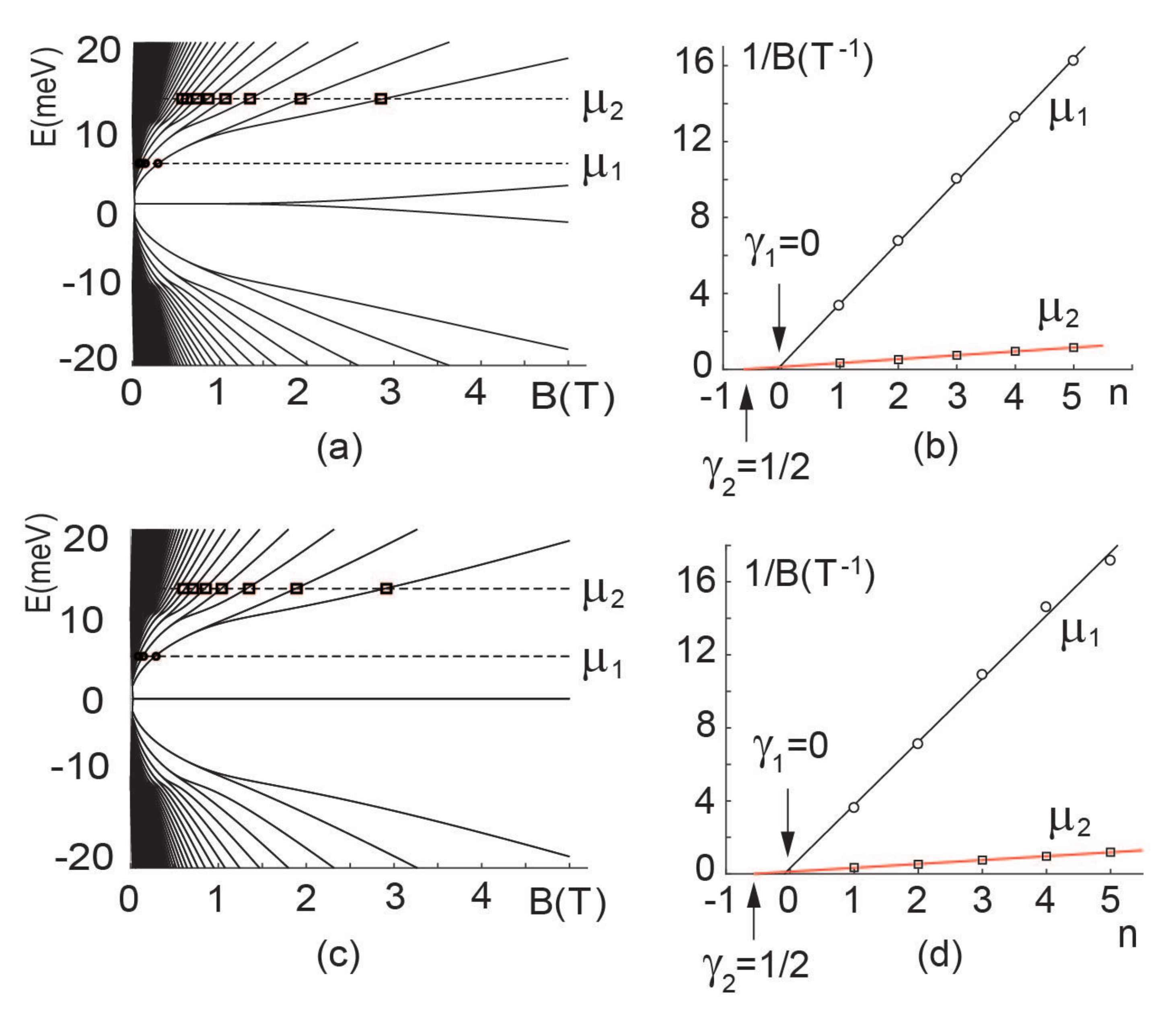}
\end{center}
\caption{(a) and (c) Parallel field numerically obtained Landau levels (full lines) of the equi-energy nodal line at $p_z=0$ for $H$ and $H_{\tr{vx}}$, respectively: there is splitting in the $n=0$ level of $H$ but not for $H_{\tr{vx}}$. (b) and (d) Landau fan diagrams with two chemical potential values $\mu_{1-2}$ for $H$ and $H_{\tr{vx}}$, respectively.}
\end{figure}

\subsection{Parallel field direction (orbit II)}
We pick the field in the $x$-direction inducing the two extremal orbits lying on the $p_x=0$ plane. The corresponding LLs are numerically obtained for $H$ (Fig. 4a) and $H_{\tr{vx}}$ (Fig. 4c). Their low-energy features can be readily understood \cite{Mon09} by making an expansion in the Hamiltonian around the band touching points $H(0,\pm p^{0}+p_y,p_z) \approx \pm (p_0/m_r)  p_y \sigma_x+ v_z p_z \sigma_y$, giving two copies (``valleys") of anisotropic Dirac Hamiltonians. Similarly as for the vortex-ring Hamiltonian, we have $H_{\tr{vx}}(0,\pm p^{0}+p_y,p_z)\approx \mp (p_0/m_\bot) p_z \sigma_y\pm (p_0/m_r) p_y \sigma_z$. For $H$, the low-energy LL spectrum is $E_n=\pm v_F\,\sqrt{n e B\hbar}$ where $v_F =\sqrt{2 p_0 v_z/m_r}$ (for $H_{\tr{vx}}$, $v_F =\sqrt{2p_0^2/m_\bot m_r}$), in agreement with the numerically obtained values for small B-field and for small energies (see Appendix). The sublevel splitting seen at higher magnetic field/energy is due to the ``valley" degeneracy lifting of the two-Dirac-cone problem \cite{Mon09}. 

A fan diagram analysis in Figs. 4b and 4c shows that below the critical chemical potential, the single slope (for $\mu_1$) corresponds to the two orbits of equal area and locks at the same increasing value as $\mu$ increases, due to the circular symmetry of the Fermi surface (orbit II in Fig.~2a). Moreover, the orbits exhibit a $\pi$ Berry phase. This is because the orbit circulating the nodal line pick up a Berry phase, as exemplified in the underlying degenerate low-energy Dirac spectrum at low field and energy. Above the critical value, the orbit encircles two Dirac points resulting in a zero and $2 \pi$ Berry phase (winding) for $H$ and $H_{\textrm{vx}}$, respectively. However, the differences in the total winding cannot be distinguished in quantum oscillations. This is due to the fact that $\gamma$ is defined only modulo unity.
 
There, however, is a distinguishing topological characteristic of the two Hamiltonians in the $n=0$ LL at high magnetic field \cite{Gail11, Gail12a, Gail12b}. For $H$, we have two Dirac Hamiltonians at low energy with the \textit{opposite} chirality whereas for $H_{\textrm{vx}}$ they display the \textit{same} chirality. When the chiralities are opposite, the $n=0$ low-energy LL wavefunctions are $(|0\rangle,0)^\textrm{T}$ and $(0,|0\rangle)^\textrm{T}$ in the occupation basis, respectively; on the other hand, when their chirality are the same, the LL wavefunctions for the two valleys are identical $(|0\rangle,0)^\textrm{T}$. When the magnetic field is sufficiently large such that the two valleys `feel' the presence of each other, the two degenerate $n=0$ wavefunctions mix and result in energy level degeneracy lifting in $H$ (Fig.~4a). However, the energy degeneracy lifting does not occur in $H_{\textrm{vx}}$ since the two valleys share the same eigenfunction, i.e., the $n=0$ eigenenergy remains pinned at $E=0$ even at high field (Fig.~4c). While this difference does not show up in quantum oscillations, it is useful to resort to probes which are sensitive to the LL structure, i.e., spectroscopic probes.

\section{Landau level results for tilted-energy nodal line}
We solve for the LL for the tilted-energy nodal line Hamiltonian $H_{\tr{tilt}}$ with the ordinary nodal line in three field directions, corresponding to the three high symmetry axes, inducing three kinds of extremal orbits I, II, III shown in Fig.~2b. With the effect of the energy tilting clarified, the result for the vortex-ring case follows similarly.  

\subsection{Perpendicular field (orbit I)}
The LLs with $p_z=0$ are given analytically by 
\beq
E_n=\pm (n+1/2-c_1-c_2^2)\epsilon_B,
\eeq  
where $n=0,1,2, \dots$, $c_1=\Delta/\epsilon_B$ and $c_2=v \sqrt{e B \hbar}/(\sqrt{2}\epsilon_B)$ is related to the strength of the tilt as characterized by $v$ (Fig.~5a and Appendix). For quantum oscillations, the results are very similar to the corresponding equi-energy nodal line case (Sec. IV A), yielding two frequencies corresponding to the two closed orbits of unequal area. The differences are that the orbits, as shown in Fig.~5c, encircle both electron and hole pockets. However, there is no signature in the quantum oscillation period when the hole pocket diminishes as $\mu$ increases. Now the Fermi surface becomes a genus 1 electron pocket, and the behavior recovers that of the equi-energy nodal line case. A Landau fan analysis shows that there is no Berry phase associated with the orbits (Fig.~5b).

\subsection{Parallel field in $y$-direction (orbit II)}
The LLs are obtained numerically with $p_y=0$ (Fig.~6a). The low-energy features can be captured by making an expansion in the Hamiltonian around $H_{\tr{tilt}}(\pm p_0+p_x,0,p_z)\approx \pm v (p_0+p_x)\,\mathbb{I} \pm  (p_0/m_r) p_x \sigma_x+ v_z p_z \sigma_y$. We see that the effects of the energy tilt are two folds. It shifts the reference energy of the two Dirac Hamiltonians away from each other, and it induces an \textit{electric} field for the two copies of anisotropic Dirac Hamiltonian. Using the analytic solution for graphene in a crossed electric-magnetic field, from Ref.~\cite{Lukose07, Peres07} (see Appendix), the low-energy LL spectrum is 
\beq
E_n=\pm v p_0\pm f(E_A,E_B) \sqrt{n},
\eeq
$n=0,1,2, \dots$, which shows a characteristic $\pi$-Berry phase of Dirac fermions (see Appendix for definitions of $f$, $E_A$, $E_B$). 

From the numerical full LL spectrum, below the critical $\mu$, we extract two sets of $\{B_n\}$ giving two slopes in the fan diagram corresponding to the sectional cuts of the electron and hole pockets, which are generally unequal in area. With $\gamma=0$, the two orbits give a $\pi$ Berry phase, as expected for orbits encircling the nodal line. Beyond the critical $\mu$, the Berry phase vanishes, in accordance with an orbit encircling the outer region of the nodal line.

\begin{figure}
\begin{center}
\includegraphics[width=8.7cm]{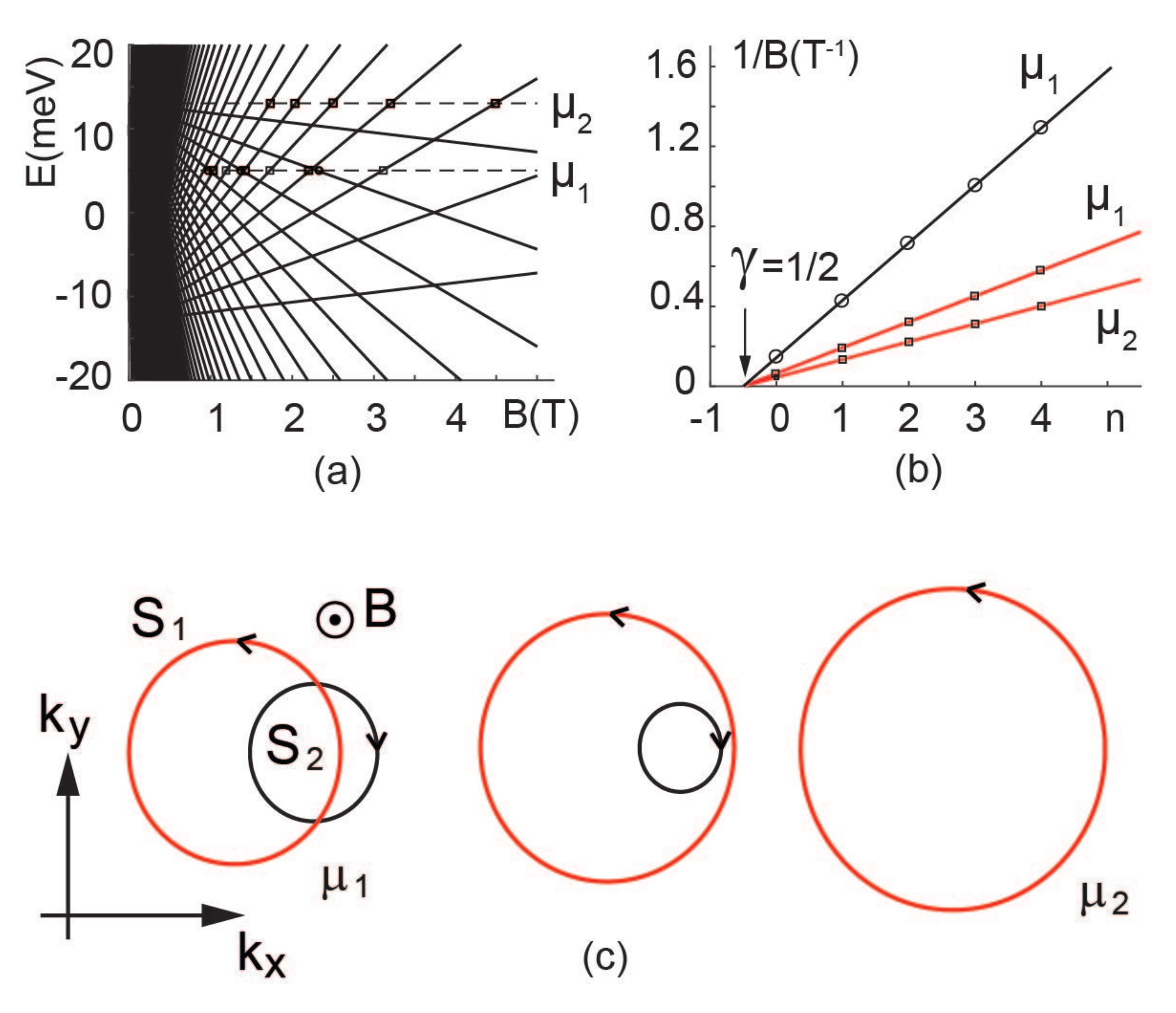}
\end{center}
\caption{(a) Landau levels of the tilted-energy nodal line for a perpendicular magnetic field at $p_z=0$. (b) Landau fan diagram with two sectors (open circles (black orbit) and open squares (red orbit)). (c) Extremal orbits for various chemical potentials.}
\end{figure}
\subsection{Parallel field in $x$-direction (orbit III)}
The extremal orbit assumes a fixed $p_x=p^0_x < 0 $ resulting in $H_{\tr{tilt}}=v p_x^0 \,\mathbb{I}+(1/2m_r)(p_y^2+(p_x^0)^2-p_0^2)\sigma_x+v_z p_z \sigma_y$. Except for an overall shift in the energy, the Hamiltonian is of the form studied in Sec. IV B, featuring two Dirac Hamiltonians in the $(p_y,p_z)$ plane for $p_0^2-(k_x^0)^2<0$. The numerically obtained LLs is shown in Fig. 7a.  At and away from $\mu=0$, we find one oscillation frequency corresponding to the extremal orbit encircling the outer region of the nodal line with no Berry phase. Of course, there is another extremal orbit with a different $p_x^0>0$ encircling the hole pocket, giving similar physical consequences.

\begin{figure}
\begin{center}
\includegraphics[width=8.7cm]{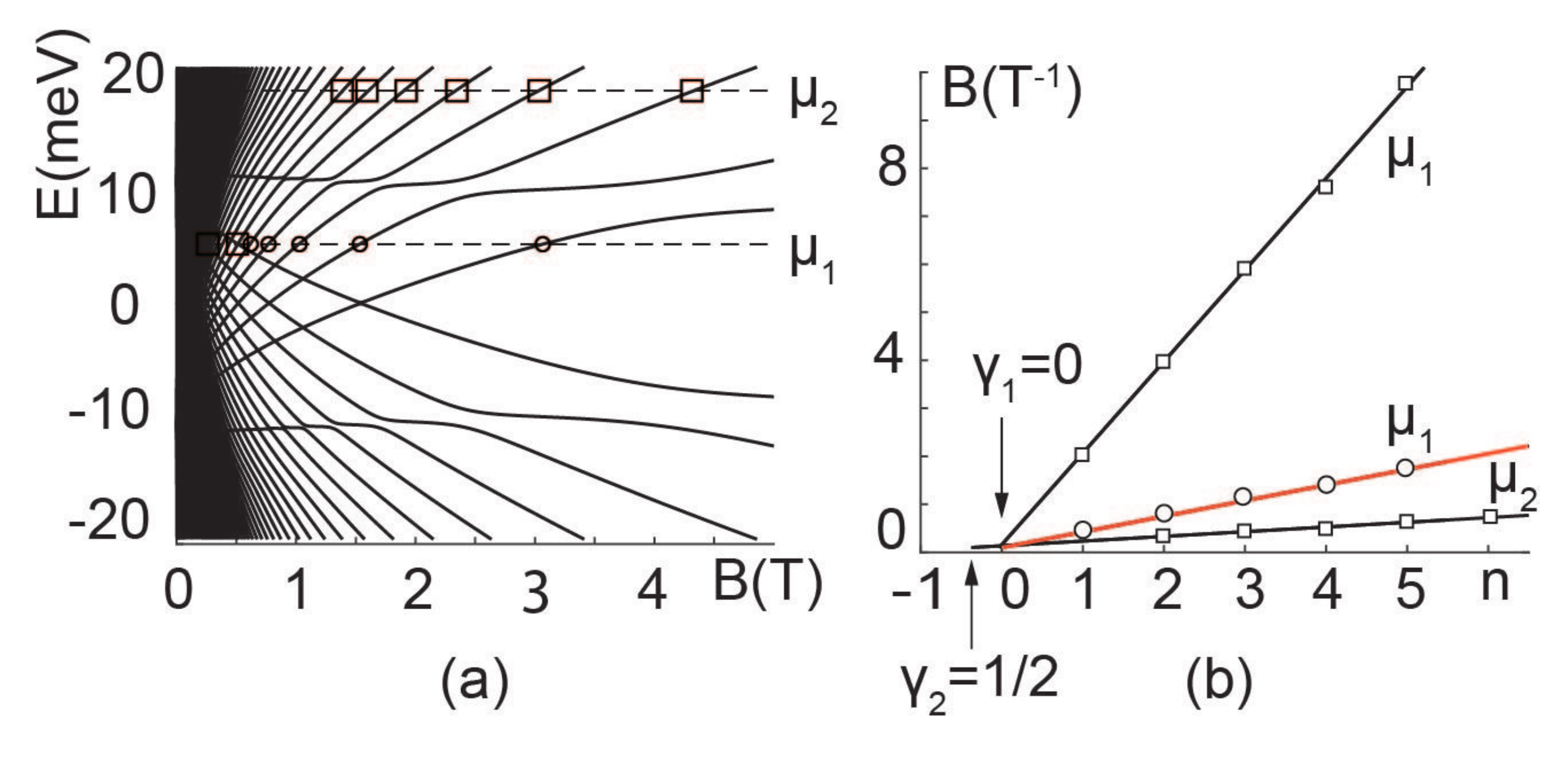}
\end{center}
\caption{(a) Numerically obtained Landau levels of the tilted-energy nodal line for a parallel $y$-direction magnetic field at $p_y=0$. (b) Landau fan diagram. $\pi$ (No) Berry phase for $\mu_1$ ($\mu_2$), respectively.}
\end{figure} 

\begin{figure}
\begin{center}
\includegraphics[width=9.cm]{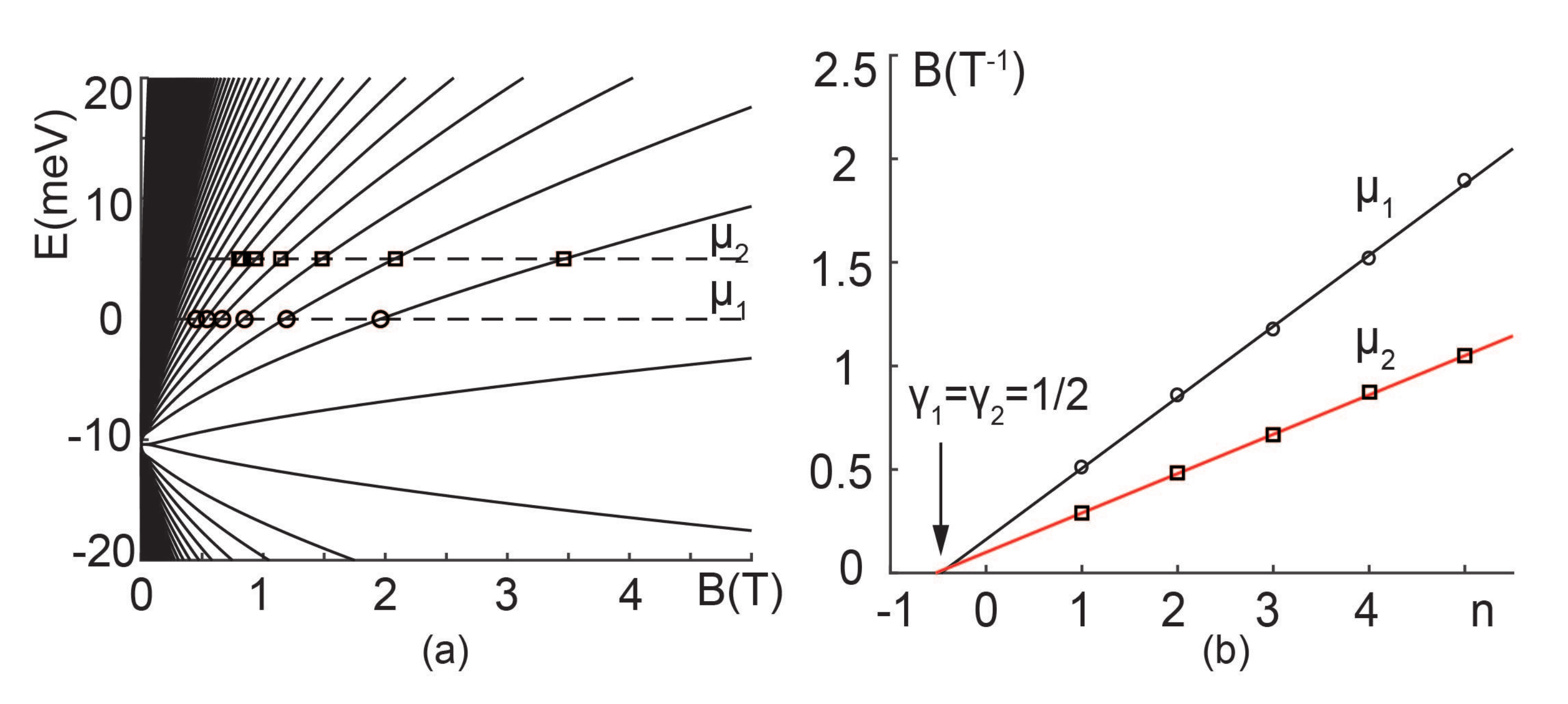}
\end{center}
\caption{(a) Numerically obtained Landau levels of the tilted-energy nodal line for a parallel $x$-direction magnetic field at $p_x^2=0.9 \,p_0^2$. (b) Landau fan diagram yielding no Berry phase.}
\end{figure} 

\section{Conclusion and outlook}
We have studied zero-temperature magnetic quantum oscillations of nodal line semimetals associated with two kinds of pseudospin textures, in the equi-energy and the tilted-energy settings, respectively. By analyzing the Landau levels structure of the extremal orbits, we find a correspondence between different parts of the spectrum and oscillation periods associated with different orbits. We extract their Berry phase signatures via Landau-fan analyses. They yield a $\pi$ Berry phase when the cyclotron orbit encircles the nodal line. In comparing the ordinary and pseudospin vortex-ring nodal lines, their distinguishing feature in the lowest Landau level, however, is not revealed in quantum oscillations.

For future work, the effect of magnetic breakdown for nearby orbits on nodal line Fermi surface with electron/hole pockets is an interesting direction. Moreover, the relation between the over-tilted-energy nodal line case (equivalent to an extreme electric field strength) and the phenomenon of Landau level collapse \cite{Lukose07, Peres07} deserves further investigations.  

We thanks Chen Fang, Long Zhang, Gilles Montambaux, and Zheng Liu for helpful discussions. This work was in part supported by Tsinghua University Initiative Research Programme, the 1000 Youth Young Fellowship China (L.-K. L.) and DFG under grant SFB 1143 (R. M.). 

\appendix

\begin{figure}
\begin{center}
\includegraphics[width=6.5cm]{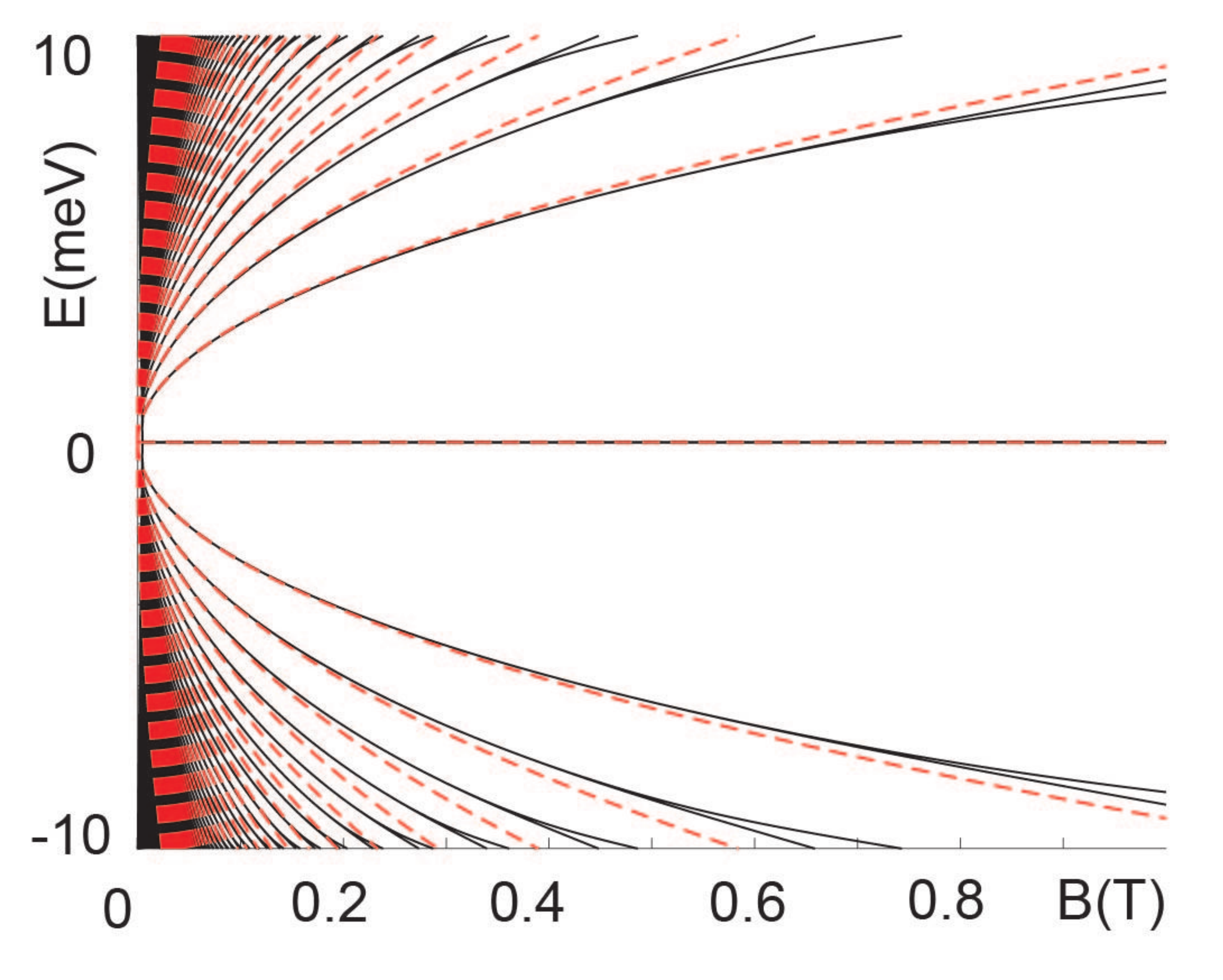}
\end{center}
\caption{The Landau levels of $H$ for a parallel field at small field. The black lines are the analytical result and the red dashed lines are the numerical result.}
\end{figure}
\section{Analytic low-energy Landau levels under different magnetic field directions}
\subsection{Equi-energy nodal line}
For an anisotropic Dirac Hamiltonian
\beq
H= v_y p_y \sigma_x + v_z p_z \sigma_y,
\eeq
an isotropic Dirac Hamiltonian is obtained after a coordinate space rescaling $y\rightarrow (v_y/v_z) y$, giving LLs
\beq
E_n=\pm v_F \,\sqrt{n e B \hbar},
\eeq
$n=0,1,2, \dots$, with an effective Fermi velocity $\sqrt{2 v_y v_z}$. Fig. 8 shows the agreement between the low-energy LL (red dashed lines) with the full numerical solution (black full lines) for a magnetic field in the parallel $x$-direction (Sec. IV B).

\subsection{Tilted-energy nodal line}
For the perpendicular field case, the Hamiltonian for the extremal orbit (with $p_z=0$ and $\sigma_x \rightarrow \sigma_z$) is
\beq
H_{\tr{tilt}}=v p_x \,\mathbb{I}+ (1/2m_r)(p_x^2+p_y^2-p_0^2)\sigma_z.
\eeq
In the presence of a magnetic field, the momentum operators become $(\hat{p}_x, \hat{p}_y)\rightarrow(\hat{p}_x-e \hat{A}_x, \hat{p}_y-e \hat{A}_y)$ with $\mathbf{A}=(-B\hat{y}/2, B\hat{x}/2,0)$. Introducing creation and annihilation operators we get
\beq
H_{\tr{tilt}}/\epsilon_B=ic_2(\hat{a}^\dag-\hat{a})\,\mathbb{I}+(\hat{a}^\dag \hat{a}+\frac{1}{2}-c_1)\sigma_z
\eeq
with $c_1=\Delta/\epsilon_B$, $c_2=v \sqrt{e B \hbar}/(\sqrt{2}\epsilon_B)$. The eigenvalue equations are 
\beq
\left(\hat{a}^\dag \hat{a}+\frac{1}{2}-c_1+ic_2(\hat{a}^\dag-\hat{a})\right)|\psi_A\rangle&=&(E_n/\epsilon_B)|\psi_A\rangle\nn\\
\left(-(\hat{a}^\dag \hat{a}+\frac{1}{2}-c_1)+ic_2(\hat{a}^\dag-\hat{a})\right)|\psi_B\rangle&=&(E_n/\epsilon_B)|\psi_B\rangle.\nn
\eeq
Defining $\hat{a}=\hat{\gamma}-i c_2 \,\,(\hat{a}^\dag=\hat{\gamma}^\dag+i c_2)$, the eigen-energies are given by Eq. (5) in Sec. V A.

For a parallel field in the $y$-direction, the low-energy Hamiltonian for the extremal orbit is given by
\beq
H_{\tr{tilt}}=\pm v p_0 \,\mathbb{I}+\begin{pmatrix}
vp_x & \pm\frac{p_0}{m_r}p_x-iv_z p_z\\
\pm\frac{p_0}{m_r} p_x+iv_zp_z & v p_x
\end{pmatrix}.
\eeq
Under a magnetic field, 
\beq
H_{\tr{tilt}}=\pm v p_0 \,\mathbb{I} +\begin{pmatrix}
E_A(\hat{a}+\hat{a}^\dag) & E_B\hat{a}^\dag\\
E_B\hat{a} & E_A(\hat{a}+\hat{a}^\dag)
\end{pmatrix}.
\eeq
Here $E_A=v \sqrt{(v_z eB\hbar/2 v_x)}$, $E_B=\sqrt{2 v_x v_z e B\hbar }$, $v_x\equiv p_0/m_r$.
Using the result from Ref.~\cite{Peres07} for graphene in a crossed electric-magnetic field, the LLs are given by (Sec. V B)
\beq
E_n=\pm v p_0\pm\frac{(E_B^2-4E_A^2)^{3/4}}{E_B^{1/2}}\sqrt{n},
\eeq
$n=0,1,2, \dots$. Fig. 9 shows the comparison between the low-energy LL (red dashed lines) and the full numerical solution (black full lines) of Sec. V B. 
 \begin{figure}
\begin{center}
\includegraphics[width=8.5cm]{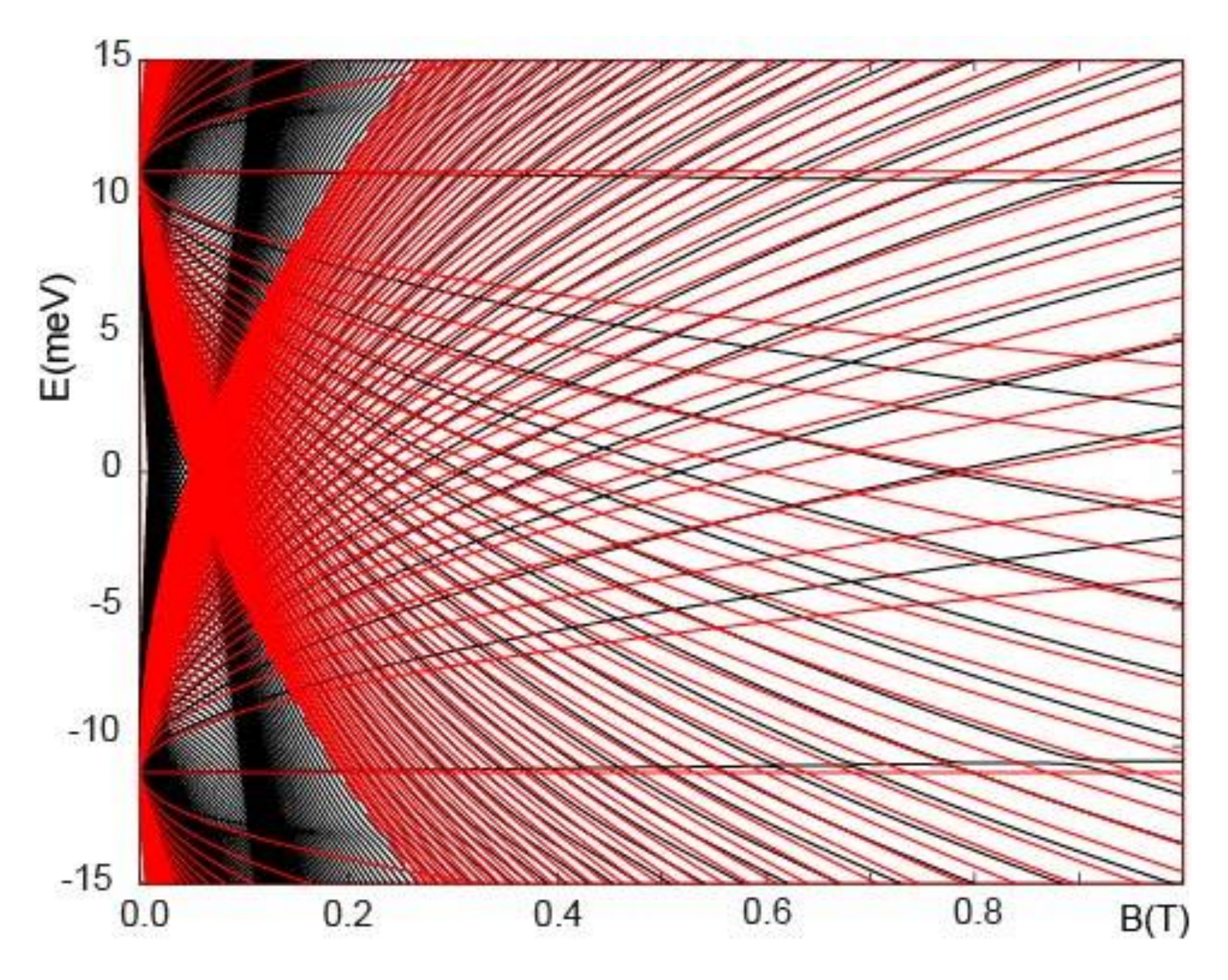}
\end{center}
\caption{The Landau levels of $H_{\tr{tilt}}$ for a parallel $y$-direction field at small field. The black lines are the analytical result and the red dashed lines are the numerical result.}
\end{figure}


\begin{thebibliography}{99}
\bibitem{ft2} Parameters chosen are typical of three-dimensional semimetals: the energy gap at the origin $\Delta\equiv p_0^2/(2 m_r)=10\tr{meV}$, the effective nodal line band mass $m_r=0.05 m_e$. The other mass is set to be isotropic $m_\bot=m_r$ for simplicity, and the Fermi velocities are chosen such that the Fermi surface volumes of $H$ and $H_{\tr{vx}}$ for finite $E$ are near identical: the $z$-direction Fermi velocity $m_r v_z^2/2=0.854\Delta$ and the tilting velocity $m_r v^2/2=0.3\Delta$.
\end{thebibliography}
\end{document}